\documentclass[aps,prl,twocolumn]{revtex4}%
\usepackage{graphics}
\usepackage{amsmath}
\usepackage{graphicx}
\usepackage{amsfonts}
\usepackage{amssymb}
\usepackage{float}
\usepackage{longtable}
\usepackage{epsfig}
\usepackage{latexsym}
\usepackage{theorem}
\usepackage{bbm}
\usepackage{bm}
\usepackage{psfrag}%
\providecommand{\U}[1]{\protect\rule{.1in}{.1in}}

\begin{document}
\title{Optimal squeezing for quantum target detection}
\author{Gaetana Spedalieri and Stefano Pirandola}
\affiliation{Department of Computer Science, University of York, York YO10 5GH, UK}

\begin{abstract}
It is not clear if the performance of a quantum lidar or radar, without an
idler and only using Gaussian resources, could exceed the performance of a
semiclassical setup based on coherent states and homodyne detection. Here we
prove this is indeed the case by showing that an idler-free squeezed-based
setup can beat this semiclassical benchmark. More generally, we show that
probes whose displacement and squeezing are jointly optimized can strictly
outperform coherent states with the same mean number of input photons for both
the problems of quantum illumination and reading.

\end{abstract}
\maketitle

\subsection{\label{Sec1}Introduction}

Quantum hypothesis testing~\cite{Hesltrom,QSD1,QSD2,Janos} is one of the most
important theoretical areas at the basis of quantum information
science~\cite{Nilelsen}. In the bosonic setting~\cite{reviewSENSING}, some of
the basic protocols are those of quantum
illumination~\cite{qill1,qill2,qill3,qill4,qill5,qill6,qill7,qill8,qill9,qill10,qill11,Quntao,Karsa}%
, aimed at better detecting the presence of a remote target in conditions of
bright thermal noise, and quantum reading~\cite{qreading}, aimed at boosting
data retrieval from an optical digital memory. These protocols can be modelled
as problems of quantum channel discrimination where quantum resources are able
to outperform classical strategies in detecting different amounts of channel loss.

One of the basic benchmark which is typically considered in assessing the
quality of quantum illumination is the use of coherent states and homodyne
detection. This is considered the best known (semi-)classical strategy and is
often adopted to assess the advantage of quantum resources (e.g.,
entanglement)~\cite{Quntao,Karsa} for lidar/radar
applications~\cite{radarBOOK,Marcum,albersheim1981}. This classical strategy
is clearly based on Gaussian resources (i.e., Gaussian states and measurement)
and does not involve any idler system. An open question is to determine if
there is another idler-free strategy based on Gaussian resources which
strictly outperforms the classical one.

In this work we answer this question positively, showing the advantage of
using displaced-squeezed states with suitably optimized amount of squeezing.
Such optimal probes are able to outperform coherent state for the same number
of mean signal photons per mode irradiated over the unknown target. While this
can be shown for quantum illumination, i.e., quantum lidar applications, the
advantage becomes more evident and useful in a different regime of parameters,
as typical for quantum reading.

\subsection{Optimized probes for target detection}

Consider the detection of a target in terms of a binary test: The null
hypothesis $H_{0}$ corresponds to target absent, while the alternative
hypothesis $H_{1}$ corresponds to target present. These hypotheses correspond
to the following quantum channels acting on a single-mode input state probing
the target:

\begin{description}
\item[H$_{0}:$] A completely thermalizing channel, i.e., a channel replacing
the input state with a thermal environment state with $\bar{n}_{B}$\ mean photons.

\item[H$_{1}:$] A thermal-loss channel with loss $1-\eta$, so that only a
fraction $\eta$\ of the signal photons survives, while $\bar{n}_{B}$ mean
thermal photons are added to the state.
\end{description}

\noindent Both channels can be represented by a beam splitter with
transmissivity $\eta$\ and input thermal noise $\bar{n}_{B}^{\prime}:=\bar
{n}_{B}/(1-\eta)$. We have $\eta=0$ for\ $H_{0}$ and some $\eta>0$
for\ $H_{1}$. In terms of quadratures $\hat{x}=(\hat{q},\hat{p})^{T}$ the
action of the beam splitter is $\hat{x}\rightarrow\sqrt{\eta}\hat{x}%
+\sqrt{1-\eta}\hat{x}_{B}$, where $\hat{x}_{B}$ is a background mode with
$\bar{n}_{B}^{\prime}$ mean photons.

As long as there is a different amount loss between the two channels above, it
is possible to perfectly discriminate between the two hypotheses if we are
allowed to use input states with arbitrary energy. However, if we assume that
the input states must have a mean number of photons equal to $\bar{n}_{S}$,
then there is an error associated with the discrimination problem.

Consider a displaced squeezed-state at the input of the unknown channel.
Assume that this state has $\bar{n}_{A}$ photons associated with its amplitude
$\alpha$, namely, $\bar{n}_{A}=\left\vert \alpha\right\vert ^{2}$. Without
losing generality, assume that $\alpha\in\mathbb{R}$, so the mean value of the
state is equal to $\bar{x}=(\sqrt{2\bar{n}_{A}},0)^{T}$ (see~\cite{Notation}%
\ for details on notation). The state has covariance matrix (CM)
$\mathbf{V}=(1/2)\mathrm{diag}(r,r^{-1})$ for position squeezing $r\leq1$
($=1$ corresponding to a coherent state). It is easy to compute that the mean
number of photons generated by the squeezing is equal to $f_{r}=(r+r^{-1}%
-2)/4$. Thus, the mean total number of photons associated with the state is
$\bar{n}_{S}=\bar{n}_{A}+f_{r}$. Note that, for fixed value of $\bar{n}_{S}$,
the amount of squeezing is bounded within the range $r_{-}\leq r\leq1$,
where~\cite{noteB}
\begin{equation}
r_{-}:=2\bar{n}_{S}+1-2\sqrt{\bar{n}_{S}(\bar{n}_{S}+1)}.
\end{equation}

Assume that the state is homodyned in the $\hat{q}$-quadrature (position). The
outcome $q$ will be distributed according to a Gaussian distribution with mean
value
\begin{equation}
\bar{q}=\sqrt{2(\bar{n}_{S}-f_{r})}\geq0,
\end{equation}
and variance $\sigma^{2}=r/2$. If homodyne is performed after the unknown
beam-splitter channel, then we need to consider the transformations%
\begin{equation}
\bar{q}\rightarrow\sqrt{\eta}\bar{q},~\sigma^{2}\rightarrow\lambda_{\eta}%
^{2}:=\frac{2\bar{n}_{B}+1-\eta(1-r)}{2}.
\end{equation}
By measuring the $\hat{q}$-quadrature for $M$ times and adding the outcomes,
the total variable $z$ will be distributed according to a Gaussian
distribution $P_{\eta}(z)$ with mean value $\bar{z}:=M\sqrt{\eta}\bar
{q}=M\sqrt{2\eta(\bar{n}_{S}-f_{r})}$ and variance $\sigma_{z}^{2}%
:=M\lambda_{\eta}^{2}$. Note that, for $H_{0}$, we have a Gaussian $P_{0}(z)$
centered in $\bar{z}=0$ with variance $\sigma_{z}^{2}=M\lambda_{0}%
^{2}=(M/2)(2\bar{n}_{B}+1)$. For $H_{1}$, we have instead $P_{1}(z)=P_{\eta
}(z)$\ with $\eta>0$.

Let us adopt a maximum likelihood test with some threshold value $t>0$
(implicitly optimized), where we select $H_{1}$ if $z>t$ (otherwise we select
the null hypothesis $H_{0}$). The false-alarm probability $p_{\text{FA}}$ and
the mis-detection probability $p_{\text{MD}}$ are therefore given
by~\cite{NoteIntegral}%
\begin{align}
p_{\text{FA}}  &  :=\mathrm{prob}(H_{1}|H_{0})=\int_{t}^{+\infty}P_{0}(z)dz\\
&  =\frac{1}{2}\left\{  1-\operatorname{erf}\left[  \tfrac{t}{\sqrt{M(2\bar
{n}_{B}+1)}}\right]  \right\}  ,\nonumber\\
p_{\text{MD}}  &  :=\mathrm{prob}(H_{0}|H_{1})=\int_{-\infty}^{t}P_{1}(z)dz\\
&  =\frac{1}{2}\left\{  1+\operatorname{erf}\left[  \frac{t-M\sqrt{2\eta
(\bar{n}_{S}-f_{r})}}{\sqrt{M[2\bar{n}_{B}+1-\eta(1-r)]}}\right]  \right\}
.\nonumber
\end{align}
For equal priors $\mathrm{prob}(H_{0})=\mathrm{prob}(H_{1})=1/2$, the mean
error probability is given by $p_{\text{err}}=(p_{\text{FA}}+p_{\text{MD}})/2$.

It is clear that the performance of the displaced squeezed states is at least
as good as that of the coherent states, because the optimization over the
squeezing parameter $r$ (within the constraint imposed by $\bar{n}_{S}$)
includes the point $r=1$. The goal is therefore to show that some amount of
squeezing can be useful to strictly outperform the coherent-state probes. For
this purpose, the first step is to correctly quantify the amount of thermal
noise $\bar{n}_{B}$ that is seen by a free-space lidar receiver.

Consider a receiver with aperture radius $a_{R}$, angular field of view
$\Omega_{\text{fov}}$ (in steradians), detector bandwidth $W$ and spectral
filter $\Delta\lambda$ (the latter can be very small thanks to the
interferometric effects occurring at the homodyne detector). Compactly, we may
define the photon collection parameter $\Gamma_{R}:=\Delta\lambda W^{-1}%
\Omega_{\text{fov}}a_{R}^{2}$ (see Ref.~\cite{FSpaper} for more details).
Considering that sky brightness at $\lambda=800~$nm is $B_{\lambda
}^{\text{sky}}\simeq1.5\times10^{-1}$ W m$^{-2}$ nm$^{-1}$ sr$^{-1}%
$~\cite{Miao,BrussSAT} (in cloudy conditions), the mean number of thermal
photons per mode hitting the receiver is
\begin{equation}
\bar{n}_{B}=\frac{\pi\lambda}{hc}B_{\lambda}^{\text{sky}}\Gamma_{R}~.
\label{downT}%
\end{equation}
Assuming $a_{R}=10~$cm, $\Omega_{\text{fov}}\simeq3\times10^{-6}$~sr
($\Omega_{\text{fov}}^{1/2}=1/10~$degree), $W=100~$MHz and $\Delta
\lambda=10^{-4}~$nm, we get $\bar{n}_{B}\simeq5.8\times10^{-2}$ mean thermal
photons per mode.

\begin{figure}[t]
\vspace{0.2cm}
\par
\begin{center}
\includegraphics[width=0.40\textwidth] {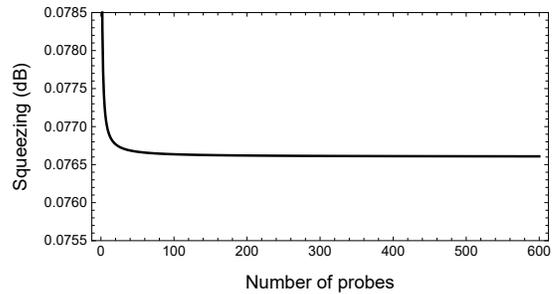}
\end{center}
\par
\vspace{-0.5cm}\caption{Optimal squeezing $-10\log_{10}r$\ versus number of
probes/modes $M$ for the problem of target discrimination.\ Parameters are
$\eta=0.2$ for target present (otherwise $\eta=0$), $\bar{n}_{S}=0.1$ mean
photons per signal mode, and $\bar{n}_{B}\simeq5.8\times10^{-2}$ mean thermal
photons per background mode. The threshold value $t$ is implicitly optimized
for each point.}%
\label{SqueezingParPic}%
\end{figure}

Let us take $\bar{n}_{S}=0.1$ signal photons per mode and assume $\eta=0.2$
for the reflectivity of the target (the latter quantity implies either a
proximity of the target or very good reflectivity properties, i.e., very
limited diffraction at the target). For realistic values of $M\lesssim10^{3}%
$~\cite{radarBOOK,Marcum,albersheim1981}, we can see that the optimal probes
are not coherent states but rather states that are both displaced and
squeezed. For the regime of parameters considered, the difference is small but
still very significative from a conceptual point of view. As we can see in
Fig.~\ref{SqueezingParPic}, the amount of squeezing is small, i.e., less than
0.08~dB. (See~\cite{Codes} for the mathematica files associated with this
manuscript.) 

The significance of the result relies on the fact that the use of coherent
states and homodyne detection might be considered to be the optimal Gaussian
strategy for quantum illumination in the absence of idlers. This is not
exactly true. One can find regimes of parameters where the presence of
squeezing can strictly outperform coherent states, even if the advantage can
be very small. As we discuss below, the difference becomes more appreciable in
problems of quantum reading~\cite{qreading} or short-range quantum
scanning~\cite{qscanning}, where the transmissivities associated with the
hypotheses are relatively high.

\subsection{Optimized probes for quantum reading or scanning}

Note that the probabilities $p_{\text{FA}}$ and $p_{\text{MD}}$ discussed
above can be extended to the general case where $P_{0}(z)=P_{\eta_{0}}(z)$ and
$P_{1}(z)=P_{\eta_{1}}(z)$ for arbitrary $0\leq\eta_{0}\leq\eta_{1}\leq1$. In
such a case, we just write $p_{\text{FA}}=\tfrac{1}{2}(1-\Omega_{0})$ and
$p_{\text{MD}}=\tfrac{1}{2}(1+\Omega_{1})$, where we define (for $u=0,1$)%
\begin{equation}
\Omega_{u}=\operatorname{erf}\left(  \frac{t-M\sqrt{2\eta_{u}(\bar{n}%
_{S}-f_{r})}}{\sqrt{M[2\bar{n}_{B}+1-\eta_{u}(1-r)]}}\right)  .
\end{equation}
This scenario can refer to the readout of an optical cell with two different
reflectivities~\cite{qreading}, or to the scan of a biological sample to
distinguish between a blank from a contaminated sample~\cite{qscanning}%
.\begin{figure}[t]
\vspace{0.2cm}
\par
\begin{center}
\includegraphics[width=0.40\textwidth] {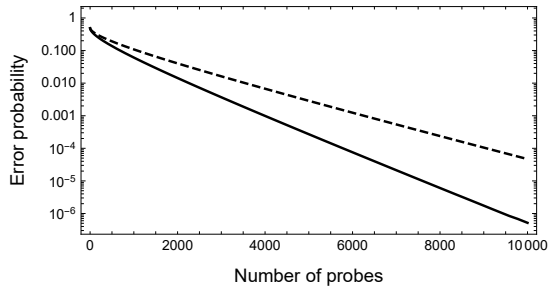}
\end{center}
\par
\vspace{-0.5cm}\caption{Optimal displaced-squeezed probes for quantum reading
and scanning. We plot the mean error probability achievable with the optimal
displaced-squeezed probes (solid) with respect to just-displaced probes, i.e.,
coherent states (dashed).\ Parameters are $\eta_{0}=0.9$, $\eta_{1}=0.98$,
$\bar{n}_{S}=1$ mean photons per signal mode, and $\bar{n}_{B}\simeq
5.8\times10^{-2}$ mean thermal photons per background mode. The squeezing
parameter $r$ and the threshold value $t$ are implicitly optimized for each
point.}%
\label{scanPIC}%
\end{figure}

For our numerical investigation, we consider high transmissivities $\eta
_{0}=0.9$ and $\eta_{1}=0.98$, and relatively-high signal energy $\bar{n}%
_{S}=1$. The other parameters are the same as above for target detection.
Thus, we study the performance for equal-prior symmetric hypothesis testing,
plotting the mean error probability $p_{\text{err}}$ as a function of the
number of probes $M$. As we can see from Fig.~\ref{scanPIC}, the optimized
displaced-squeezed probes (here corresponding to about $4~$dB of squeezing)
clearly outperform coherent states with orders of magnitude advantage for
increasing $M$.

\begin{figure}[t]
\vspace{0.2cm}
\par
\begin{center}
\includegraphics[width=0.40\textwidth] {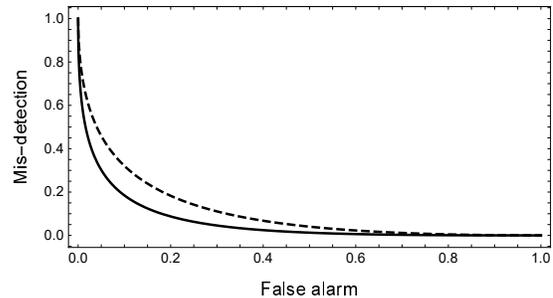}
\end{center}
\par
\vspace{-0.5cm}\caption{Receiver operating characteristic (ROC) $p_{\text{MD}%
}$ as a function of $p_{\text{FA}}$. We compare the performance of the optimal
displaced-squeezed probes (solid) with respect to just-displaced probes, i.e.,
coherent states (dashed).\ Parameters are $M=500$, $\eta_{0}=0.9$, $\eta
_{1}=0.98$, $\bar{n}_{S}=1$ mean photons per signal mode, and $\bar{n}%
_{B}\simeq5.8\times10^{-2}$ mean thermal photons per background mode. The
squeezing parameter $r$ is implicitly optimized for each point.}%
\label{scanROCpic}%
\end{figure}

We also consider asymmetric hypothesis testing~\cite{GaeGAUSS,Kli,Quntao}
plotting the receiver operating characteristic (ROC), expressed by the
misdetection probability versus the false-alarm probability for some fixed
number of probes. As we can see from Fig.~\ref{scanROCpic}, for the case of
$M=500$, we have a clear advantage of the optimized probes with respect to
coherent states. This behaviour is generic and holds for other values of
$M$.

\subsection{Conclusions}

In this work we have investigated the use of displaced-squeezed probes for
problems of bosonic loss discrimination, i.e., quantum illumination and
quantum reading. We have compared the performance of these probes with respect
to that of purely-displaced ones, i.e., coherent states, showing that a strict
advantage can be obtained by opimizing over the amount of squeezing while
keeping the input mean number of photons as a constant. For the specific case
of target detection, our results show that there exists an idler-free
Gaussian-based detection strategy outperforming the typical (semi-)classical
benchmark considered in the literature, which is based on coherent states and
homodyne detection. Due to the intrinsic Gaussian nature of the process, the
dependence of the quantum advantage versus the various parameters is
continuous and expected to be maintained in the presence of small experimental
imperfections of the devices.

\bigskip

Acknowledgements.--~This work was funded by the EU Horizon 2020 Research and
Innovation Action under grant agreement No. 862644 (FET Open project: Quantum
readout techniques and technologies, QUARTET). We thank the anonymous referee
for spotting a numerical instability in our previous figure 1 and providing associated feedback on our Mathematica code.


\begin{thebibliography}{99}
\bibitem {Notation}We work in the standard quantum optics notation, so the
annihilation operator reads $\hat{a}=(\hat{q}+i\hat{p})/\sqrt{2}$ in terms of
the quadratures $\hat{q}$ (position) and $\hat{p}$ (momentum), with $[\hat
{q},\hat{p}]=i$. This means that the amplitude reads $\alpha=(q+ip)/\sqrt{2}$,
so the mean value is given by $\bar{x}=(q,p)^{T}=\sqrt{2}(\operatorname{Re}%
\alpha,\operatorname{Im}\alpha)^{T}$. Then, the number operator takes the form
$\hat{n}:=\hat{a}^{\dag}\hat{a}=(\hat{q}^{2}+\hat{p}^{2}-1)/2$, implying that
a displaced and $q$- (or $p$-) squeezed state has mean number of photons
$\bar{n}=(\left\langle \hat{q}^{2}\right\rangle +\left\langle \hat{p}%
^{2}\right\rangle -1)/2=|\alpha|^{2}+[\mathrm{var}(\hat{q})+\mathrm{var}%
(\hat{p})-1]/2$. Finally, note that the quadrature variance of a thermal state
corresponds to $\bar{n}+1/2$, where $\bar{n}$ is the mean number of photons
and $1/2$ is the vacuum shot noise.
\bibitem {noteB}Note that, in general, we write $r_{-}\leq r\leq$ $r_{+}$,
where $r_{+}:=2\bar{n}_{S}+1+2\sqrt{\bar{n}_{S}(\bar{n}_{S}+1)}\geq1$.
However, since we consider position-squeezing in our analysis, we implicitly
have $r\leq1$, so the upper bound $r_{+}$ is just replaced by $1$.
\bibitem {NoteIntegral}Recall that, for a Gaussian $g(z)=(\sigma\sqrt{2\pi
})^{-1}\exp[-(z-\bar{z})^{2}/(2\sigma^{2})]$ we have $\int_{t}^{+\infty
}g(z)dz=(1-\Omega)/2$ and $\int_{-\infty}^{t}g(z)dz=(1+\Omega)/2$ where
$\Omega:=\operatorname{erf}[(t-\bar{z})/(\sigma\sqrt{2})]$.
\end{thebibliography}
\end{document}